\documentclass[aps,prfluids,amsmath,amssymb,preprint]{revtex4-2}

\usepackage{graphicx}  
\graphicspath{{Figures/}}

\usepackage{xcolor}

\newif\ifhighlight
\highlighttrue 

\newcommand{\hl}[1]{\ifhighlight\textcolor{black}{#1}\else#1\fi}
\usepackage{lineno}
\usepackage[hypertexnames=false]{hyperref}  

\usepackage{cancel}

\date{\today}

\begin{abstract}
When a slice of beet is placed on a plate with a thin layer of beet juice, one can observe a clear fringe around the beet, where the color is more translucent than the rest of the juice. 
The hypotheses in literature were inconsistent and limited, which motivated us to revisit this phenomenon. 
Using a motorized confocal displacement sensor, we measured the  temporal evolution of the liquid surface profile across the fringe. 
Our findings suggest that a suction flow, induced by the capillary rise of the contact line, causes a dimple – a small concave depression – to form on the liquid surface. 
While surface tension and gravity tends to smooth out the dimple, viscous drag acts against them if the liquid film is sufficiently thin. 
Our scaling analysis correctly estimates the dependence of dimple lifetime on liquid properties and film thickness. 
We also capture the dimple formation dynamics by numerically solving the lubrication equation with the Young-Laplace equation. 
This work provides a new interpretation for a common phenomenon.
\end{abstract}

\begin{document}

\title{Fringe around a Beet Slice: Wetting-induced Dimple in a Thin Liquid Film}

\author{Zhengyang Liu}
\affiliation{Department of Biological and Environmental Engineering, Cornell University, Ithaca, NY 14853, USA}
\author{Yicong Fu}
\author{Abhradeep Maitra}
\author{Kunal Kumar}
\affiliation{Sibley School of Mechanical and Aerospace Engineering, Cornell University, Ithaca, NY 14853, USA}
\author{Justin Chen}
\author{Sunghwan Jung} \thanks{sj737@cornell.edu}
\affiliation{Department of Biological and Environmental Engineering, Cornell University, Ithaca, NY 14853, USA}

\maketitle

\section{Introduction}

\begin{figure}
    \centering
    \includegraphics[width=1\linewidth]{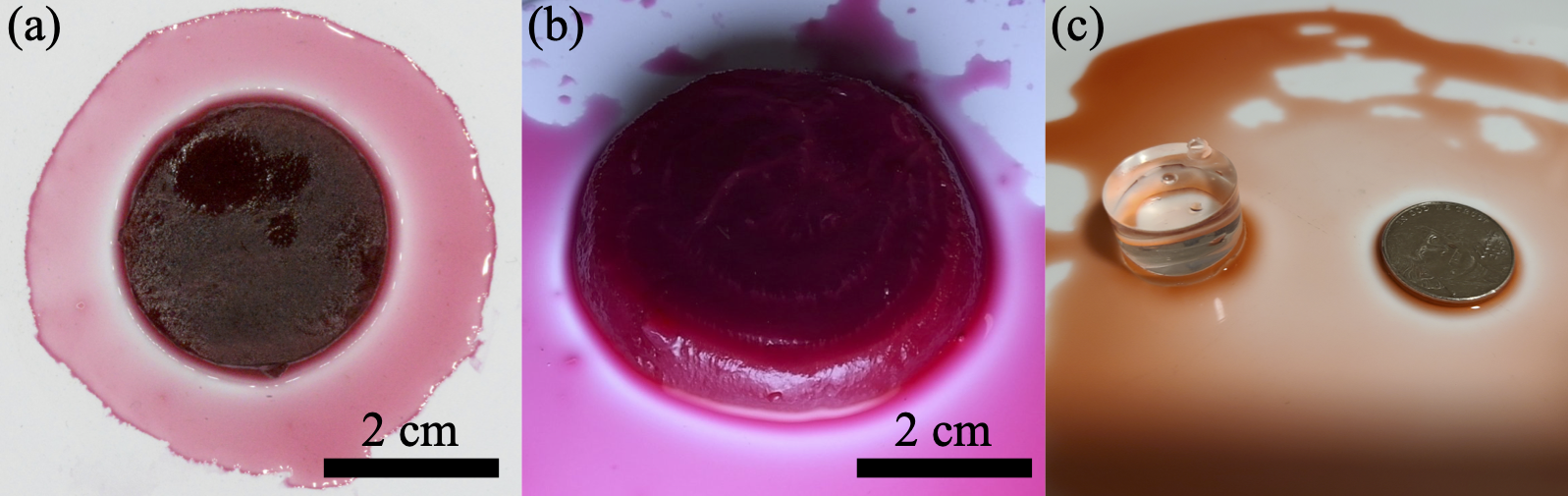}
    \caption{
    The fringe around solid objects.
    (a) Beet top view.
    (b) Beet oblique angle view.
    (c) A polydimethylsiloxane (PDMS) cylinder and a coin.
    }
    \label{fig:beet-fringe-observation}
\end{figure}

If a slice of beet is placed in a thin layer of juice, a clear fringe will be observed around the beet, the color of which is more translucent than the rest of the juice film (Figs.~\ref{fig:beet-fringe-observation}(a-b) and Supplemental Movie S1). 
This phenomenon is commonly observed in kitchens and has been published as a curious observation \cite{Satterly1956}. 
Two hypotheses, based on the absorption of coloring matter and the deformation of the liquid surface, have been proposed to explain this phenomenon \cite{Satterly1956,Scott1982}. 
The fact that these two hypotheses are inconsistent and that both have limitations motivate us to revisit this phenomenon. 
The first observer, Satterly, presumed that there was a ``suction'' effect which drained the thin liquid film \cite{Satterly1956}. 
However, two questions remain open: (i) how does the suction effect arise? and (ii) how does the suction generate the fringe pattern?

The suction could arise from two possible mechanisms: porousness and wetting. 
Beet root, like all other plant roots, has many vascular tissues that enable water and nutrient transport \cite{Eshel2013}.
This makes the beet root a porous material, which absorbs liquid.
Although it is intuitive to attribute the suction effect to the porousness of the beet root, we observe that the suction effect arises even if the beet root is saturated with liquid. 
Actually, even a non-porous material, such as a coin, can also induce the fringe pattern (Fig.~\ref{fig:beet-fringe-observation}(c)).
An alternative suction mechanism is the contact line rise due to surface wetting \cite{deGennes1985,Bonn2009}.
The wettability of beet juice on a beet surface is expected to be high for two reasons. 
First, the beet is saturated with beet juice already, so the molecular affinity is presumably higher. 
Second, the beet surface appears to be rough, which induces superhydrophilicity \cite{Quere2008}.
As we will show, wetting is indeed the main cause of the suction effect. 

Regarding how the suction induces the fringe pattern, Scott attempted to unify a series of independent observations of fringe patterns, including the fringe around the beet slice observed by Satterly, into the same framework of \textit{Reynolds ridge} \cite{Scott1982}. 
Reynolds ridge, a phenomenon first described by Osborne Reynolds, arises when a surfactant layer is compressed by a flowing stream underneath it against a barrier \cite{reynolds1901papers}.  
The formation of Reynolds ridge has been verified in a number of experiments with the minimal set of ingredients, namely a flowing stream, some surfactant and a barrier \cite{Sellin1968, McCutchen1970, Scott1982}, and a theoretical model has been formulated based on the picture \cite{Harper1974}.
Although the visual effect is in qualitative agreement with the fringe pattern around a beet slice, as both appear to be a ``line'' or a ``band'', the fluid flows in the two cases are quite different. 
Moreover, Reynolds ridge requires constant flow underneath the surface, the suction flow around a beet slice only lasts for a short period of time (from particle tracking velocimetry measurement, as shown in Appendix~\ref{app:flow} Fig.~\ref{fig:flow-measurement}), much shorter than the lifetime of the fringe pattern. 
Moreover, faster flow leads to a closer distance between the Reynolds ridge and the barrier \cite{Mockros1968}. 
Such dependence, however, is not observed for the fringe around the beet slice.
We, therefore, believe that this is a misinterpretation.

\begin{figure}
    \centering
    \includegraphics[width=1\linewidth]{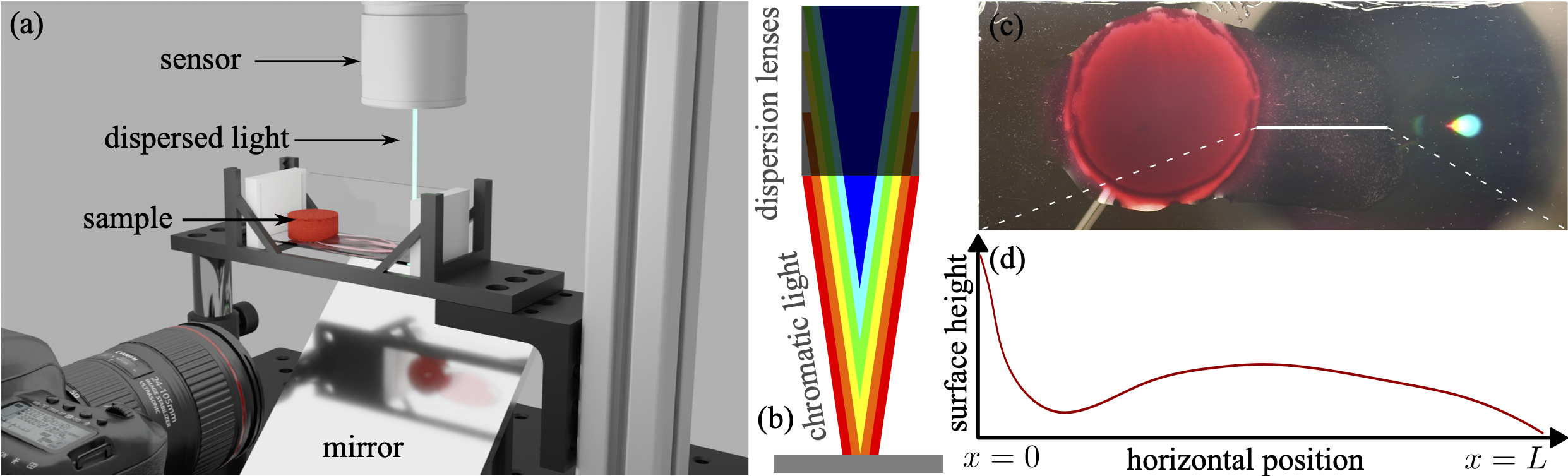}
    \caption{
    Experimental setup and techniques.
    (a) The experimental setup, where the confocal displacement sensor is mounted on a motorized linear stage. A camera is used to image the flow in the liquid simultaneously.
    (b) Illustration of the working principle of the confocal displacement sensor.
    (c) Bottom view of the sample. This is a typical image taken by the camera. 
    (d) An example of liquid surface scan data on the white line in (c). 
    }
    \label{fig:thin-film-setup}
\end{figure}

In this work, to understand this fringe pattern, we employed a confocal displacement sensor (Keyence CL-P070) to measure the surface profiles directly.
Our experiment reveals that the visual fringe is caused by a dimple, i.e. a local liquid surface depression, between the beet slice and the rest of the liquid film. 
\hl{The formation and the evolution of the dimple resembles the ripple spreading in relaxing thin polymer films \cite{Cormier2012,McGraw2012,Jalaal2019} and upon the contact between a solid and a thin liquid film \cite{Garcia-Gonzalez2023}.}
Upon the initial solid-liquid contact, a suction flow is induced by the beet slice due to the rise of the contact line, causing a dimple to form. 
Then, the competition between surface tension, hydrostatic pressure, and viscous drag determines the fate of the dimple.
\hl{Our scaling analysis correctly estimates the dependence of dimple lifetime on initial film thickness.}
Our numerical model based on the lubrication equation and the Young-Laplace equation captures the dimple formation process and also explains why such fringe is only observed when the liquid film is sufficiently thin.
Taken together, our experiment, numerical and theoretical models provide a quantitative understanding of the fringe pattern.

\section{Materials and Methods}

In our experiment, a confocal displacement sensor was mounted on a motorized linear stage  (Schneider Electric), allowing automatic, precise and fast 1D surface profile scans. 
Canned pickled beet slices (GreatValue\textsuperscript{TM}) and the juice in the same can (referred to as ``beet juice'' from now on) were used as the model materials. 
To assess the effect of surface tension and viscosity, various other liquids were also tested, as listed in Table.~\ref{tab:liquids}.
The surface tension was measured with an automatic interfacial tension meter (Shanghai Fangrui Instrument) and the viscosity was measured with a vibro viscometer (A\&D Company).
A schematic of the setup is shown in Fig.~\ref{fig:thin-film-setup}(a). 
A DSLR camera (Nikon D610) was used to image the flow in the liquid film simultaneously. 
The working principle of the confocal displacement sensor is illustrated in Fig.~\ref{fig:thin-film-setup}(b). 
Basically, white light is dispersed by a lens system, and by analyzing the color of the reflection light, the distance to the reflecting surface can be obtained.
Figure~\ref{fig:thin-film-setup}(c) is a typical image taken by the camera during the surface scan, and Fig.~\ref{fig:thin-film-setup}(d) shows the surface height measurement on the white line indicated in Fig.~\ref{fig:thin-film-setup}(c). 
In all the surface profile data shown in this work, the beet is put at $x=0$.
Supplemental Movie S2 shows the bottom view video of the scanning process, along with the surface profiles.
In the scan result, a dimple can be observed right next to the beet slice. 
It is a thinner part of the liquid film lying in the middle of the meniscus and the rest of the liquid film. 
The location of this dimple is in qualitative agreement with the fringe pattern, suggesting that the dimple is the cause of the fringe pattern. 
A thinner liquid film would allow more light to transmit.
On a white surface, for instance, the thinner part would appear whiter. 

\begin{table}[htbp]
    \centering
    \begin{tabular}{ccc}    
        \hline
        Liquid & Surface tension (mN/m) & Viscosity (mPa$\cdot$s) \\
        \hline
        Water           & 72.0 & 1.0  \\
        Beet juice      & 42.0 & 10.0 \\
        60wt\% glycerol & 65.6 & 10.5 \\
        80wt\% glycerol & 63.8 & 58.0 \\
        \hline
    \end{tabular}
    \caption{Surface tension and viscosity of liquids.}
    \label{tab:liquids}
\end{table}

\section{Results}

\subsection{Surface profiles at various initial film thicknesses}

\begin{figure}
    \centering
    \includegraphics[width=.5\linewidth]{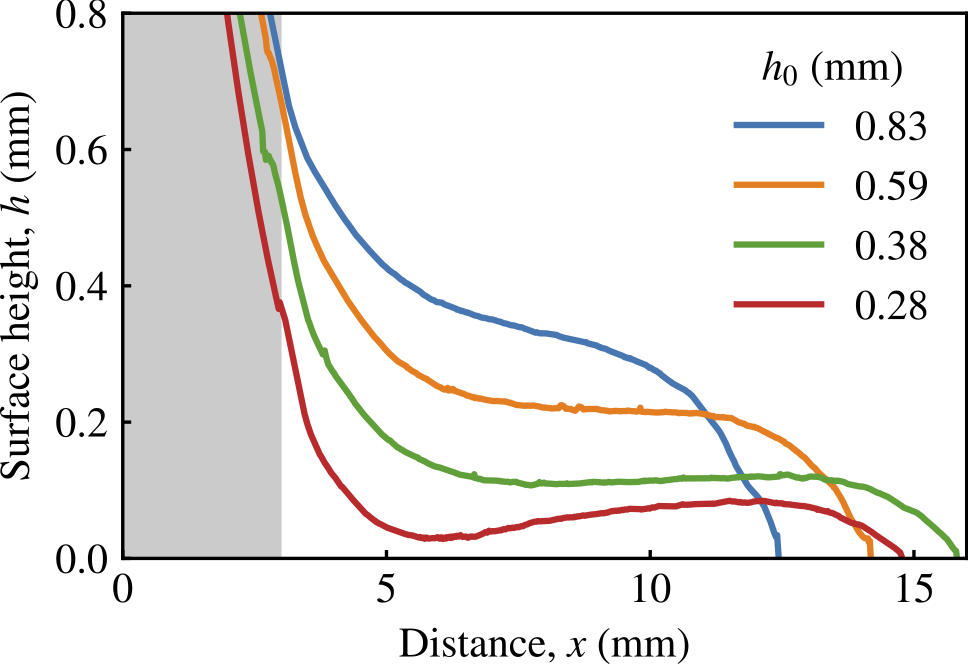}
    \caption{
    Thin film surface profiles after the contact with beet slice at $t\approx 10\;\mathrm{s}$.
    }
    \label{fig:various-initial-thickness}
\end{figure}

We prepared thin films with initial thickness $h_0$ ranging from 0.28 mm to 0.83 mm and put a slice of beet on them. 
Then, we scanned the surface profiles at $t=10$ seconds after the solid-liquid contact.
The results are shown in Fig.~\ref{fig:various-initial-thickness}.
The $x=0$ side is the interface between the liquid and the beet, and the other end of the surface profile curve is the contact line of the liquid film with the glass substrate, which is typically pinned throughout the experiment.
In close proximity to the vertical wall, the displacement sensor was unable to detect the surface height due to the large slope.
The close-to-wall (shaded in gray in the plot) part of the surface profile was constructed by extrapolation (see Appendix~\ref{app:extrapolation} Fig.~\ref{fig:extrapolation} for details of the extrapolation process).
A dimple was observed at $h_0 = 0.28$ mm.
However, for $h_0 \ge 0.28$ mm, no clear dimple was observed.
This surface profile measurement was also in qualitative agreement with our preliminary observation that the fringe only arises in sufficiently thin liquid films.

\subsection{Dimple formation process}

\begin{figure*}
    \centering
    \includegraphics[width=\linewidth]{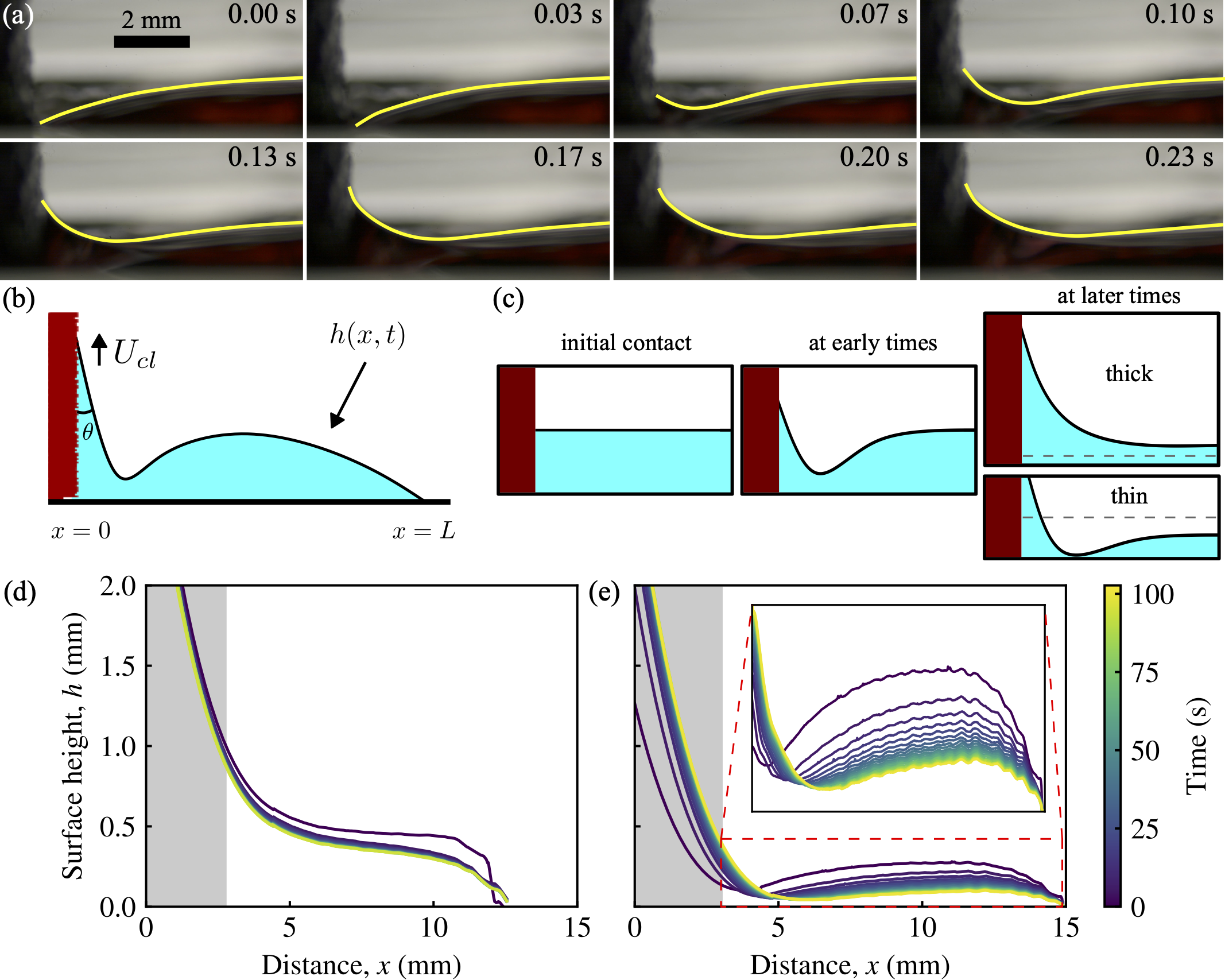}
    \caption{
    Surface profile measurement and dimple formation process.
    (a) Direct dimple formation observation.
    (b) Conceptual picture of dimple formation. A dimple forms due to the contact line rising upon the solid-liquid contact.
    (c) The fates of a dimple. In a thick film, a dimple gets smoothed out quickly, while in a thin film, a dimple lasts longer. 
    (d-e) Surface evolutions of a thick film $h_0=0.66\;\mathrm{mm}$ and a thin film $h_0=0.30\;\mathrm{mm}$. The area shaded in gray indicates the extrapolated surface profile near the meniscus.
    The inset of (e) shows the zoom-in view of the surface in the red dashed box.
    }
    \label{fig:dimple-formation-process}
\end{figure*}

The dimple formation process, i.e. the surface profile evolution during the solid-liquid contact, is crucial to understanding the fluid physics behind the dimple formation. 
Here, we first observed the evolution of the liquid surface by direct imaging, as shown in Fig.~\ref{fig:dimple-formation-process}(a).
The beet slice was gently brought into contact with a thin film of beet juice with $h_0\approx 1\;\mathrm{mm}$. 
Upon solid-liquid contact at $t=0\;\mathrm{s}$, the contact line on the out-skirts of the beet chunk quickly rose up.
A dimple was formed within 0.1 second and became smoother over time as a result of surface tension.

With direct observations, we get a rough picture of the dimple formation process, as illustrated in Fig.~\ref{fig:dimple-formation-process}(b).
First, the contact line rises very quickly at a velocity $U_{\mathrm{cl}}$ upon solid-liquid contact.
This sudden rise requires the liquid to move from the bulk thin film to the meniscus.
However, in a thin liquid film, the fluid cannot move as fast due to the viscous drag from the bottom no-slip boundary.
As a result, a dimple forms right next to the meniscus.
Once formed, the dimple has two fates depending on the film thickness: ephemeral dimples for thick films or long-standing dimples for thin films, as illustrated in Fig.~\ref{fig:dimple-formation-process}(c).

We note that direct imaging only works for liquid films on the thicker end ($h_0\ge 0.5\;\mathrm{mm}$).
For thinner films, direct imaging suffered from poor resolution.
Therefore, we used the confocal displacement sensor for thinner films, which provided much better vertical resolution.
Figures~\ref{fig:dimple-formation-process}(d) and \ref{fig:dimple-formation-process}(e) show the surface evolutions of a thick film $h_0=0.66\;\mathrm{mm}$ and a thin film $h_0=0.30\;\mathrm{mm}$ measured by the confocal displacement sensor, respectively.
No dimples were observed in the thick film and long-standing dimples were observed in the thin film, as expected.
The inset of Fig.~\ref{fig:dimple-formation-process}(e) shows the zoom-in view of the red dashed box, providing a better visualization of the formation and evolution of the dimples.

\subsection{Thin film equation and numerical simulations}

The dynamics of a thin liquid film is modeled using film drainage equation Eq.~(\ref{eq:thin-film}), also known as the Reynolds equation \cite{guyon2015physical}.
Coupled with the Young-Laplace equation Eq.~(\ref{eq:YL}), rich dynamics arising from the interactions between viscous drag, surface tension and gravity can be derived.
In 1D, the equations read
\begin{equation}\label{eq:thin-film}
    \frac{\partial h}{\partial t} = \frac{1}{3\mu} \frac{\partial}{\partial x} \left( h^3 \frac{\partial p}{\partial x} \right),
\end{equation}
\begin{equation}\label{eq:YL}
    p = -\sigma \left[1+\left(\frac{\partial h}{\partial x}\right)^2\right]^{-3/2} \frac{\partial^2 h}{\partial x^2} + \rho g h,
\end{equation}
where $h(x,t)$ is the liquid surface profile, $\mu$ is the liquid viscosity, $p(x,t)$ is the pressure in the liquid film, $\sigma$ is the surface tension, $\rho$ is the density of the liquid and $g$ is the gravitational acceleration.
Figure~\ref{fig:dimple-formation-process}(b) shows a schematic of the theoretical model.
\hl{Although it seems inconsistent to include the prefactor $[1+(\partial h/\partial x)^2]^{-3/2}$ in Eq.~(\ref{eq:YL}) with a thin film approximation, we find that this prefactor in simulations better predicts experimental surface profiles (see Fig.~\ref{fig:prefactor}). }
At the left boundary, where $x=0$, we put a wetting surface where the contact line moves at a velocity $U_{\mathrm{cl}}$ to approach the stationary contact angle $\theta_{\mathrm{s}}$.
The contact line rise is modeled by the Hoffman-de Gennes equation \cite{Chen1988,Fermigier1991,Eggers2005,Kim2017}
\begin{equation}
    U_{\mathrm{cl}} = \frac{\sigma\kappa}{\mu} \theta (\theta^2 - \theta_{\mathrm{s}}^2),
\end{equation}
where $\kappa$ is a coupling parameter and $\theta_{\mathrm{s}}$ is stationary contact angle. 
Both $\kappa$ and $\theta_{\mathrm{s}}$ can be determined by tracking contact line motion in side view images like Fig.~\ref{fig:dimple-formation-process}(a), and the results are shown in Table~\ref{tab:contact-line-dynamics}.
At the right boundary, where $x=L$, the contact line is pinned on the substrate.
At both boundaries, zero-gradient condition is applied for pressure.
The boundary conditions are summarized below:
\begin{equation}
    \frac{\partial h}{\partial t}(x=0) = U_{\mathrm{cl}},\, h(x=L)=0,
\end{equation}
\begin{equation}
    \frac{\partial p}{\partial x}(x=0) =0,\, \frac{\partial p}{\partial x}(x=L) =0.
\end{equation}
Here, no free parameter is used. 
With initial conditions where $h(x, t=0)=h_0$ is a constant, we numerically solve the equations for the thin film dynamics over time.
Finite difference method is used to discretize the spatial domain.
A method based on backward-differentiation formula (scipy BDF solver in Python \cite{2020SciPy-NMeth}) is used to integrate the equations over time. 

\begin{table}[ht]
    \centering
    \begin{tabular}{cccc}
    \hline
    Liquid          & $\kappa\times 10^{4}$ & $\theta_s$ ($^\circ$) \\
    \hline
    Beet juice      & $2.2\pm 0.2$            & $17.4\pm 6.3$           \\
    60wt\% glycerol & $1.9\pm 1.3$            & $31.2\pm 0.3$          \\
    80wt\% glycerol & $5.2\pm 1.7$            & $32.5\pm 2.6$           \\
    \hline
    \end{tabular}
    \caption{
    Contact line velocity coupling factor $\kappa$ and static contact angle $\theta_s$.
    }
    \label{tab:contact-line-dynamics}
\end{table}


\subsection{Validations}

\begin{figure*}
    \centering
    \includegraphics[width=.65\linewidth]{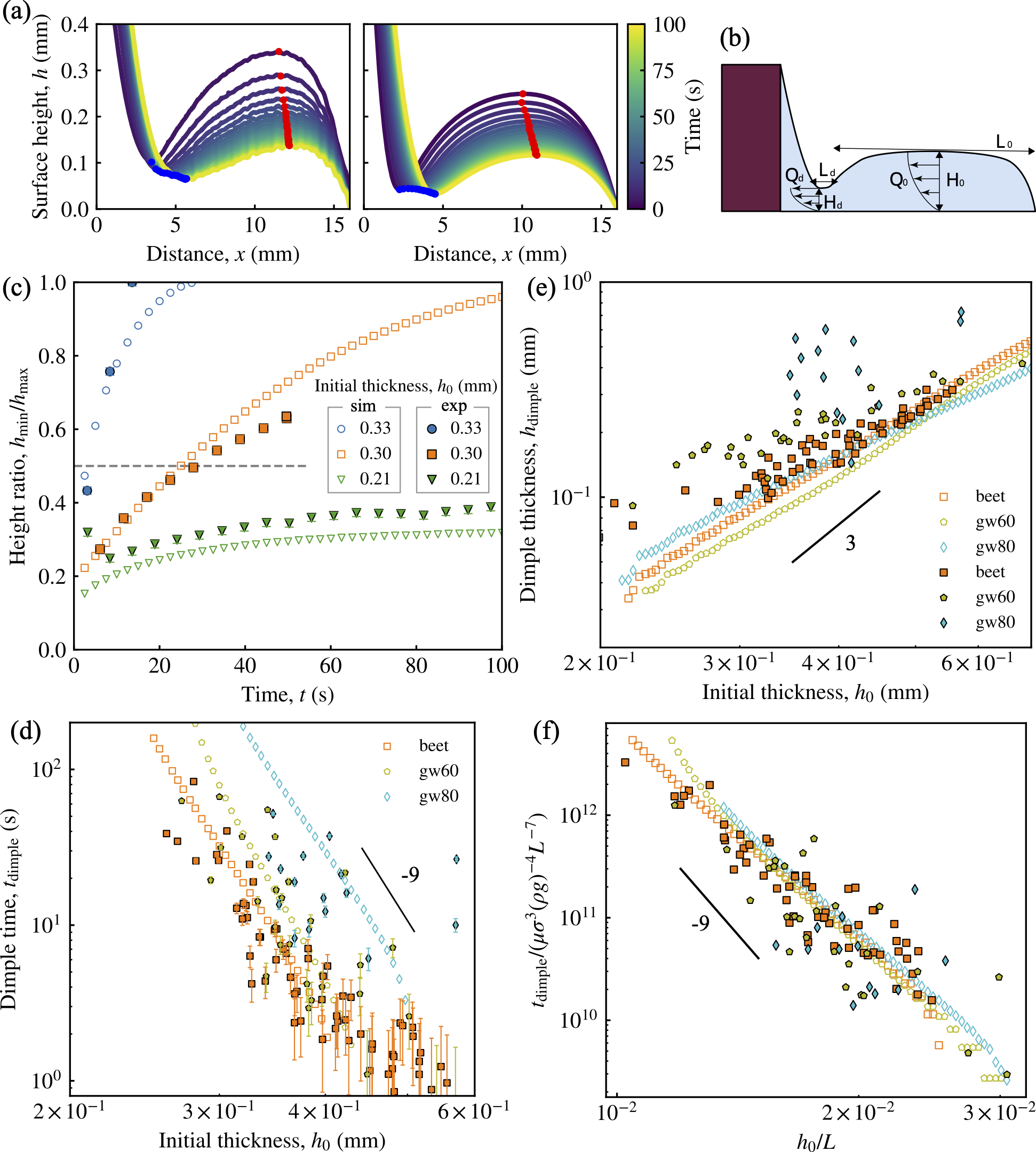}
    \caption{
    \hl{Dimple time vs. initial film thickness.} 
    (a) Compare experimental (left) and simulated (right) surface evolutions for $h_0=0.21$ mm. The dimples and film apexes of each surface profile are indicated as blue and red dots, respectively.
    \hl{The errorbars indicate the error associated with the precision of the height measurement, which is estimated to be $\pm 0.01$.}
    \hl{(b) Schematic showing two regions of interest for scaling arguments. 
    (c) Height ratio temporal evolution for $h_0=0.33,\,0.30,\,0.21\;\mathrm{mm}$ from both experiment (solid) and simulation (hollow) of beet juice. 
    (d) $t_{\mathrm{dimple}}$ at various initial film thickness $h_0\in[0.2,0.6]$ (mm), measured in various liquids, from both experiment (solid) and simulation (hollow).
    ``gw60'' and ``gw80'' stands for glycerol-water mixture with 60\% and 80\% of glycerol, respectively, by weight. 
    (e) Dimple thickness $h_\mathrm{min}$ vs. initial film thickness $h_0$. Both thicknesses are measured from the first scan in experiments and at $t=0.5$ sec for simulations.} 
    \hl{(f) Non-dimensialized $t_{\mathrm{dimple}}$ and initial film thickness. Experimental and simulation data collapse into a single curve. 
    A power-law of -9 exponent is indicated with the black triangle.
    The errorbars indicate the error associated with the surface scan process, which is estimated to be $\pm 1$ second. } }
    \label{fig:simulation}
\end{figure*}

Figure~\ref{fig:simulation}(a) compares the experimental and simulated surface profile evolutions at $h_0=0.21\;\mathrm{mm}$ (see Supplemental Movie S3 for animations comparing experimental and numerical surface profile evolutions).
\hl{
The same color code for time is used to allow comparison of the surface evolutions.
}
In both cases, we observe the formation of a sharp dimple at the beginning, followed by a gradual smooth-out. 
To quantify the lifetime of the dimple, we define dimple time $t_{\mathrm{dimple}}$ as the time it takes for the ratio between the dimple height $h_\mathrm{min}$ and the bulk film apex height $h_\mathrm{max}$ to reach 0.5. 
In both experimental and simulated data, we detect local minima and maxima in the surface profiles as indicated by blue and red dots in Fig.~\ref{fig:simulation}(a).
In Fig.~\ref{fig:simulation}(c), we plot the height ratio as a function of time at various initial thicknesses $h_0=0.21,\,0.30,\,0.33\;\mathrm{mm}$.
Regardless of the initial thickness, the height ratio increases over time.
The threshold ratio 0.5 is indicated by the gray dashed line, above which we say the dimple is ``dead''.

With this threshold, we simulated the surface evolution at $h_0\in[0.2,0.6]\;\mathrm{mm}$ for various liquids and compared it with experimental measurements.
The $t_{\mathrm{dimple}}$ results are shown in Fig.~\ref{fig:simulation}(d).
Experimental and simulated data are plotted in solid markers and hollow markers, respectively. 
For all the liquids we tested, $t_{\mathrm{dimple}}$ decreases with increasing $h_0$, agreeing with our experiments that dimples are only observed in thin films.
The simulation also reveals effects from viscosity and surface tension. 
Generally, higher viscosity leads to longer $t_{\mathrm{dimple}}$ for the same $h_0$.
This is expected because higher viscosity results in larger resistance to surface tension. 
The effect of surface tension seems counterintuitive.
According to the Young-Laplace equation Eq.~(\ref{eq:YL}), higher surface tension leads to a larger restoration pressure, which should smooth the dimple in a shorter time, i.e. smaller $t_\mathrm{dimple}$.
However, if we compare $t_\mathrm{dimple}$ of beet juice and 60\% glycerol, which have similar viscosity and different surface tension, we notice that 60\% glycerol has a higher surface tension and also a higher $t_{\mathrm{dimple}}$ at fixed $h_0$. 
Although this result contradicts with our expectation, we find it  consistent between simulations and experiments.
To understand this, we did simulations at much finer time steps to resolve the early stage dynamics for two liquids with different surface tension.
The result shows that the liquid with higher surface tension has a smaller $h_\mathrm{min}$, i.e. a deeper dimple, at early times. 
This is expected because the contact line rise is faster for the liquid with higher surface tension ($U_{\mathrm{cl}}\propto\sigma$).
The deeper dimple leads to two consequences: (i) a high surface-tension liquid has a smaller $h_\mathrm{min}/h_\mathrm{max}$ to start with, (ii) the viscous drag is stronger due to the thinner film. 
As a result, higher surface tension leads to a longer $t_{\mathrm{dimple}}$.

\subsection{The scaling relation between dimple time and initial thickness}\label{sec:time-scale}

\hl{The lubrication equation, Eq.~(\ref{eq:thin-film}), can be non-dimensionalized as 
\begin{equation}
        Ca \frac{\partial h^*}{\partial t^*} = \frac{1}{3} \left(\frac{\mathcal{H}}{\mathcal{L}}\right)^4 \frac{\partial}{\partial x^*} \left( h^{*3} \frac{\partial }{\partial x^*} \left( \frac{\partial^2 h^*}{\partial x^{*2}} + Bo \, {h^*} \right) \right) \, ,        
\end{equation}
where $\mathcal{H}$ and $\mathcal{L}$ are the characteristic height and length. Here, the capillary number is defined as $Ca = \mu V/\sigma = \mu \mathcal{H}/\sigma \mathcal{T}$ and the Bond number is defined as $Bo = \rho g \mathcal{L}^2 /\sigma$. Assuming $\mathcal{L}$ is the horizontal length of the initial film, our experiments have $\mathcal{L} \sim 24$ mm, which is larger than the capillary length $\lambda_\mathrm{cap} = (\sigma/\rho g)^{1/2}\sim 2$ mm. Then, the Bond number becomes $\mathcal{O} (100)$. 
When the Bond number is high, the time scale can be expressed as 
\begin{equation}
    \mathcal{T} \sim t_{\mathrm{dimple}} \sim \frac{\mu}{\rho g} \frac{\mathcal{L}^2}{\mathcal{H}^3} \label{eq:lub_time} \,.
\end{equation}
To further understand the film thickness dependence, we consider the flux balance in two different regions: a dimple and a thin film. The horizontal velocity is proportional to $ v_x \sim \frac{1}{\mu} \frac{\partial p}{\partial x} z (z- 2h(x))$, and the total flux is given as $Q \propto \int_0^h v_x dz$. The flux in the bulk thin film region is 
\begin{equation}
    Q_0 \sim \frac{1}{\mu} \frac{\rho g H_0}{L_0} {H_0^3} \,, 
\end{equation}
where $H_0$ and $L_0$ are the characteristic height and thickness of the bulk thin film region as illustrated in Fig.~\ref{fig:simulation}(b). 
The flux through the necking dimple region is
\begin{equation}
    Q_d \sim \frac{1}{\mu} \frac{\sigma H_d}{L_d^3} {H_d^3} \,,
\end{equation}
where $H_d$ and $L_d$ are the characteristic height and thickness of the dimple as also illustrated in Fig.~\ref{fig:simulation}(b). 
These two fluxes must be equal due to incompressibility. 
Assuming that $H_d \propto L_d$ and $H_0 \propto L_0$, one gets 
$H_d \propto H_0^3/\lambda_\mathrm{cap}^2$ where $\lambda_\mathrm{cap} = (\sigma/\rho g)^{1/2}$ is the capillary length. This relation is verified with experimental data in Fig.~\ref{fig:simulation}(e) where we plot the $h_\mathrm{min}$ vs. $h_0$ from the first scanned profile. }

\hl{Assuming the characteristic height $\mathcal{H}$ as $H_d$ in Eq.~(\ref{eq:lub_time}), and $H_0\sim h_0$, we find the characteristic time as  
\begin{equation}
    \frac{t_{\mathrm{dimple}}}{\mu \sigma^3 /(\rho g)^4 L^7} \propto \frac{1}{(h_0/L)^9} \label{eq:exponent_9} \,.
\end{equation} 
This relation between non-dimensional time scale and non-dimensional height is in good agreement with both experimental and simulated data (a best-fit power law $t_{\mathrm{dimple}}\propto h_0^{-\alpha}$, where $\alpha\approx 9.1\pm 0.7$) as in Fig. \ref{fig:simulation}(f). 
The left-hand side scales as $Bo^4 \, Ca^{-1}$, while the right-hand side depends solely on the ratio of characteristic height to length. This inverse relation between $Bo$ and $Ca$ is quite common in lubrication film dynamics \cite{pandey2023optimal}. The high-order dependence on the Bond number arises from the fluid flow through the narrowing dimple region. }

\section{Conclusion and discussions}

We reinterpret the fringe pattern formation around a beet slice as a dimple in the thin liquid film. 
Our new surface profile measurements and  theoretical model reveal unambiguously that the dimple formation is caused by the rise of the contact line, instead of the suction due to the porousness of beet.
That said, any material with good wetting property should induce similar fringe patterns in thin liquid films.
\hl{Whether the fringe can be observed or not depends on the dimple lifetime, which is governed by the interplay between surface tension, hydrostatic pressure, and viscous drag.
This competition can be described by the capillary number $Ca$ and the Bond number $Bo$.
When $Bo\ll 1$, surface tension dominates and the dimple is smoothed out very quickly, so no fringe can be observed.
When $Bo\gg 1$, hydrostatic pressure dominates and results in a long-lasting dimple. 
Our scaling model quantitatively captures the dimple lifetime as a function of the initial thickness. Our work resolved the inconsistent explanations of a commonly observed phenomenon.}

The comparison between experiments and simulations, as shown in Fig.~\ref{fig:simulation}(a), shows that the dimple formation process is well captured by our theoretical model.
In particular, the simulation captures the gradual horizontal shift of the dimple location, and the thinning of the bulk liquid film.
However, if we take a close look at the experimental and simulation results, we notice a discrepancy: the surface profile is more asymmetric in the bulk part of the experiment.
More mass seems to be distributed to the right.
In simulation, however, the surface profile looks almost symmetric. 
This can be attributed to evaporation and possibly Marangoni flow.
Evaporation is most significant at the contact line with the glass substrate on the right, driving a flow toward the right.
In addition, the evaporation of more volatile species can induce a Marangoni flow towards the right, similar to the famous tears of wine phenomenon \cite{Thomson1855}.
\hl{The asymmetry is not as pronounced in the surface profiles of glycerol-water mixtures (Appendix~\ref{app:marangoni}).
Since Marangoni flow is absent in glycerol-water mixtures, this is encouraging evidence that the Marangoni effect is responsible for the asymmetric surface profiles.}

\hl{
While our work identifies the wetting of the sidewall as the primary source of suction, one may wonder whether the beet porousness can also play a role. 
First, we note that in an experiment, the beet slices we used were pretty saturated with juice, so we did not expect the pore suction to be significant. 
We also tried to model the suction flow by considering the negative pressure caused by the pores, which drives the suction flow. The negative pressure was on the order of 1000 Pa ($\approx \sigma /r$ where $r $ is a typical pore radius; $\sim 50 \,\mu\mathrm{m}$ \cite{beetpore}). 
When running the numerical simulation with this negative pressure as the boundary condition, the solver failed because the suction was so drastic that no convergence could be reached. 
After screening a range of negative pressure values, we realized that by lowering the boundary negative pressure to the order of 10 Pa, we could get thin film behaviors very similar to the experiment. 
This way, we realize that the main driving force of the dimple/fringe formation is the capillary rising on the side wall of the beet, which indeed induces negative pressure on the order of 10 Pa. 
}

Lastly, we want to note that Fig.~\ref{fig:simulation}(d) shows relatively few experimental measurements of $t_{\mathrm{dimple}}$, compared to the number of scans we have done. 
In many trials, the liquid films were either too thick or too thin, so $t_{\mathrm{dimple}}$ was close to 0, or longer than 100 s, which is beyond the time of our experiment.  
These data were omitted in the log-log plot in Fig.~\ref{fig:simulation}(d).
In a linear scale plot, we show all our $t_{\mathrm{dimple}}$ measurements (see Appendix~\ref{app:all-dimple-time-data}).
A clear separation in the initial thickness $h_0$ can be identified, highlighting that long dimples are only observed for small $h_0$.
\hl{In terms of the choice of our experiment time, we consider the evaporation time scale. Typically, the evaporation time depends on the relative humidity ($RH$) in the lab. Under a humid condition ($RH = 80\%$), the time for the beet juice to evaporate is estimated as $\approx {\rho h_0}/[D_v {(C_s - C_\infty)}/{L_\mathrm{diffusion}}] \sim {\rho h_0^2}/{(D_vC_s(1-RH))} \simeq 800$ sec. Here, $D_v$ is the diffusion rate of water vapor in the air ($2.5\times 10^{-5}\;\mathrm{m^2/s}$), $C_s$ is the vapor concentration near the surface (saturation; $0.023\;\mathrm{kg/m^3}$), $C_\infty$ is the water vapor concentration far away, and $L_\mathrm{diffusion}$ is the diffusion length scale. Therefore, we chose the maximum duration of our experiments to 100 sec, which is approximately an order of magnitude smaller than the evaporation time ($\mathcal{O} (1000)$ sec).}

\hl{This initial study opens up several questions for future work. First, the assumptions made in our scaling arguments have yet to be experimentally validated. Second, the parameter space explored in our experiments, particularly with respect to film length, has been relatively limited. For shorter films, no fringe pattern is observed, whereas for longer films, the emergence of fringe patterns and the potential influence of three-dimensional effects might come in play. Finally, our use of the thin-film equation may break down near the contact point on the beet vertical surface. In order to accurately capture the dynamics in this region, we may require solving the full Navier–Stokes equations.}

\noindent\textbf{Competing interests}
The authors have no competing interests.

\noindent\textbf{Author Contributions}
SJ conceived the idea; ZL, KK, and SJ designed and built the experiment; ZL, JC, and KK collected and analyzed data. ZL, YF, AM, and SJ interpreted the results; All authors revised the manuscript. 

\noindent\textbf{Data Accessibility}
Data, figures, and simulation codes are available in https://osf.io/5sq27/ .

\section{Acknowledgement}

The authors would like to thank Y. Sun, Y. J. Suh, Dr. C. Roh and Dr. M. Wu for sharing lab equipments and for their help with experiments and Dr. D.A. Dillard for scientific discussions at the beginning of the project. 
We would also like to thank Dr. Y. Qiao, Dr. X. Ma and Dr. S. Kumar for fruitful discussions, and Zak Kujala for sharing fringe pattern pictures. 
This work was supported by National Science Foundation (CBET-2401507).

\appendix

\renewcommand{\thefigure}{A\arabic{figure}} 
\renewcommand{\thetable}{A\Roman{table}} 
\setcounter{figure}{0} 
\setcounter{table}{0} 

\section{Flow measurement}\label{app:flow}

Flow velocity measurement was done by tracking the motion of tracer particles.
We show that the flow velocity is decaying rapidly over time. 
In the mean time, the fringe pattern can last for a longer time.
This disproved the hyposthesis that the fringe pattern is an example of Reynolds ridge.

\begin{figure}[ht]
    \centering
    \includegraphics[width=.7\linewidth]{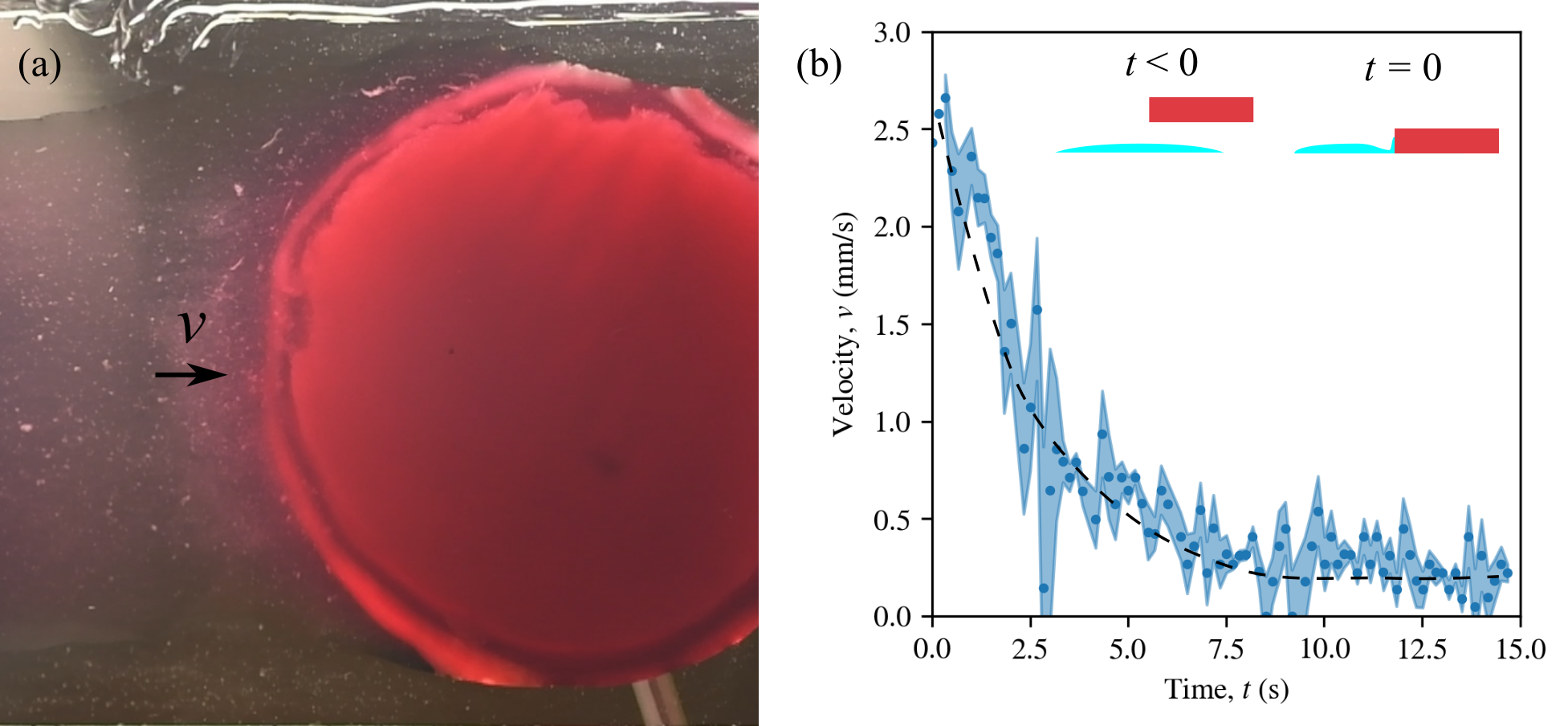}
    \caption{
    (a) A bottom view of image to measure the flow speed near the beet. The black arrow indicates the location where flow speed is measured.
    (b) Flow speed as a function of time. The insets illustrate a conceptual picture of the dimple formation process.
    }
    \label{fig:flow-measurement}
\end{figure}

\newpage

\section{Extrapolation of surface profiles}\label{app:extrapolation}

Very close to the vertical wall, the displacement sensor was not able to detect the surface height due to the large slope.
The close-to-wall part of the surface profile was constructed by extrapolation.
The gray curve is the raw data from the displacement sensor. 
The blue part of the curve is used for the extrapolation fitting. The red curve is the extrapolated surface profile. 
The high plateau on the right end of the plot is the top surface of the beet slice, which can be reliably detected by the displacement sensor. 
Our extrapolation therefore only fill in the missing part between the ``tip'' and the beet wall. 
The extrapolation is done by fitting the data with a 3rd order polynomial function.
\hl{This extrapolation is not an accurate account for the meniscus shape, and is done only for presenting the meniscus shape more realistically.
The key finding ($t_\mathrm{dimple}$) of this work does not rely on this extrapolated data.}

\begin{figure}[ht]
    \centering
    \includegraphics[width=.8\textwidth]{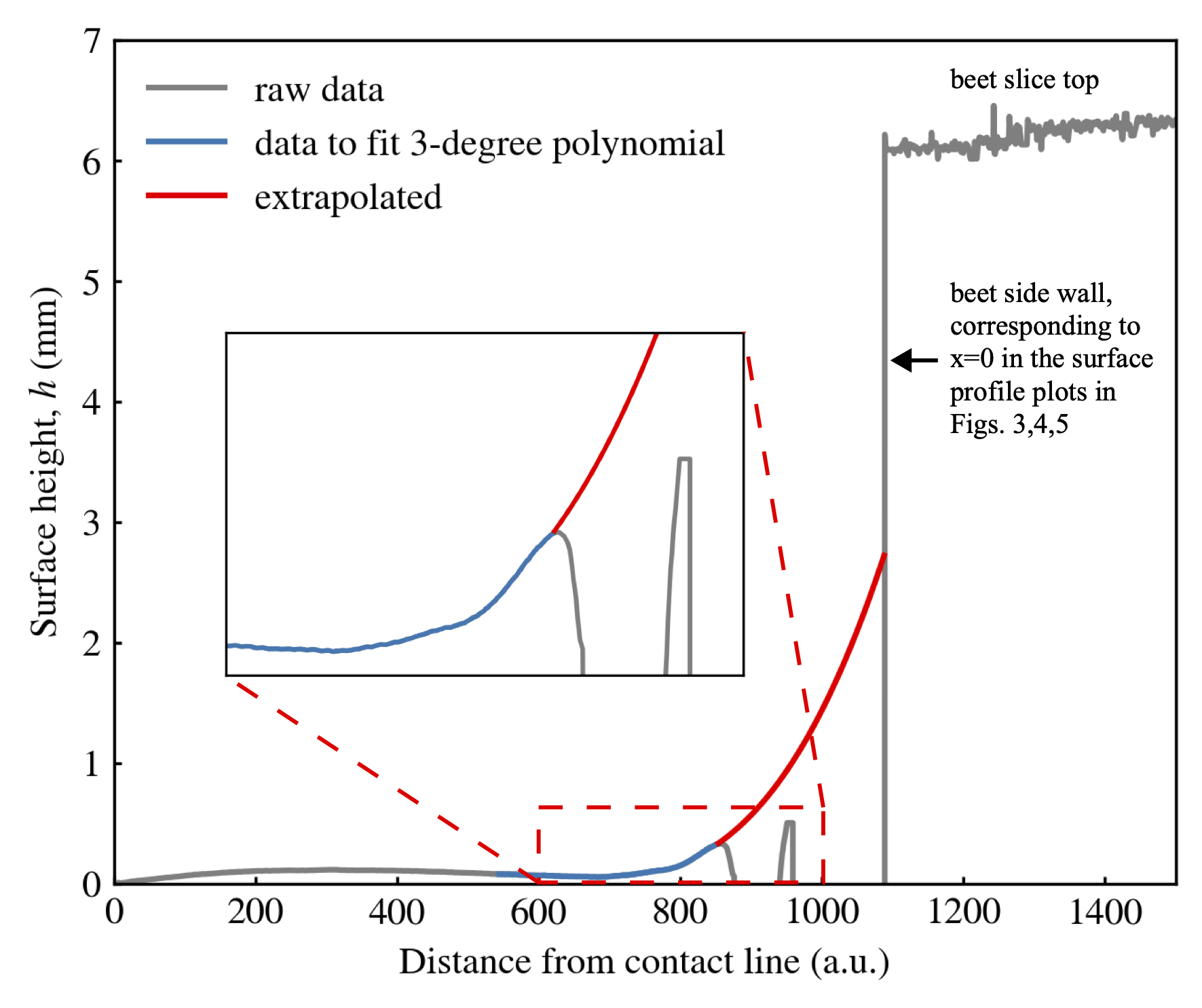}
    \caption{
    Illustration of the extrapolation method. 
    }
    \label{fig:extrapolation}
\end{figure}

\newpage

\newpage

\section{Influence of \hl{surface tension and film length}}\label{app:parameters}

\begin{figure}[ht]
    \centering
    \includegraphics[width=\textwidth]{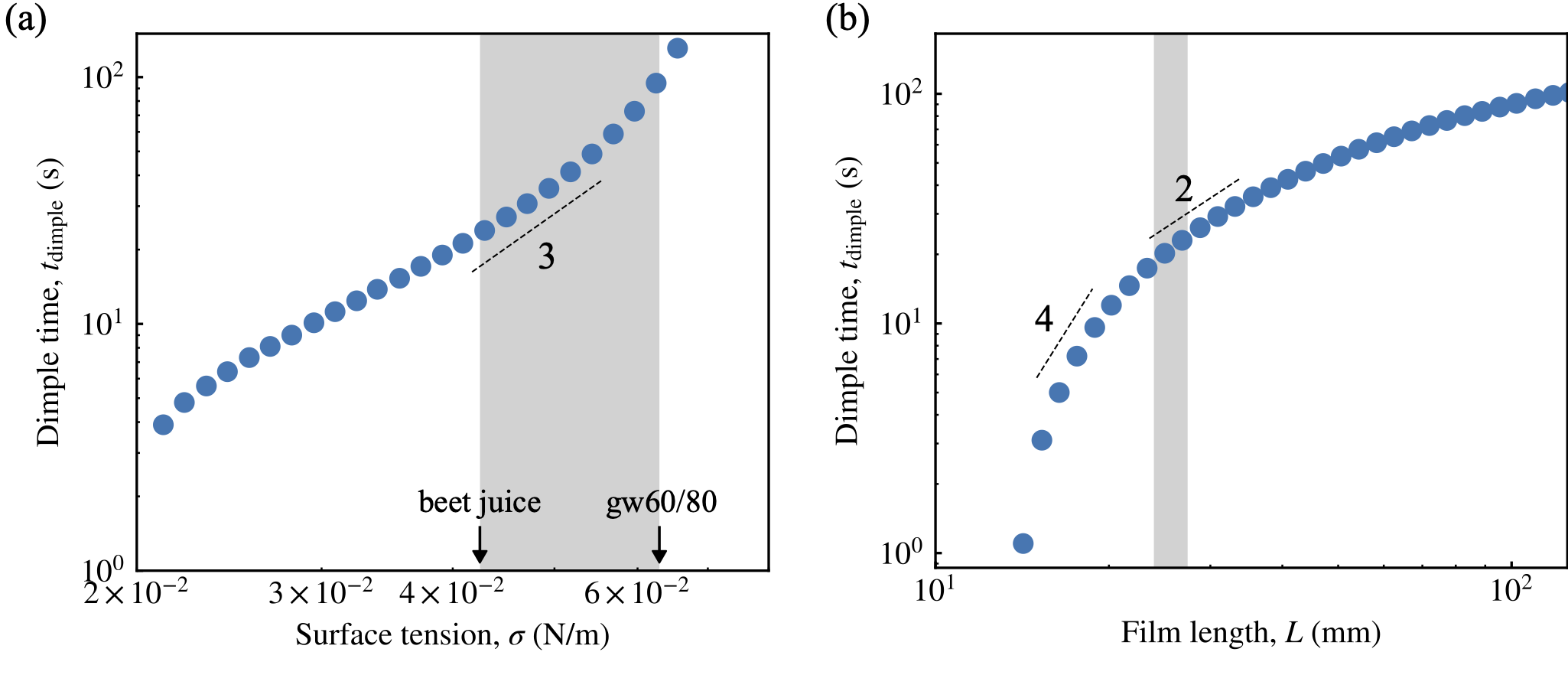}
    \caption{
    \hl{Dimple time vs. (a) surface tension or (b) film length.}
    }
    \label{fig:surfL}
\end{figure}

\section{Influence of a prefactor in curvature}\label{app:prefactor}

\begin{figure}[ht]
    \centering
    \includegraphics[width=\textwidth]{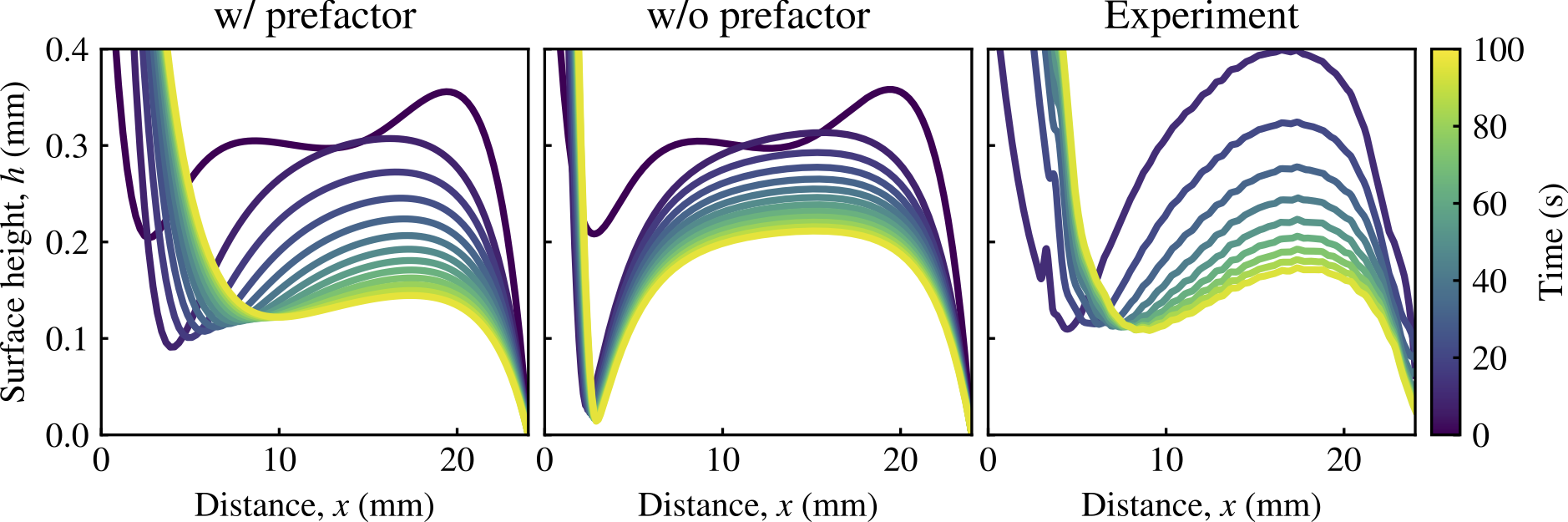}
    \caption{
    \hl{Simulated thin film surface evolutions with and without the prefactor $[1+(\partial h/\partial x)^2]^{-3/2}$ in the surface curvature, and an experimental surface evolution with the same initial condition.}
    }
    \label{fig:prefactor}
\end{figure}

\newpage

\section{Asymmetric surface profile}\label{app:marangoni}

\begin{figure}[ht]
    \centering
    \includegraphics[width=.8\linewidth]{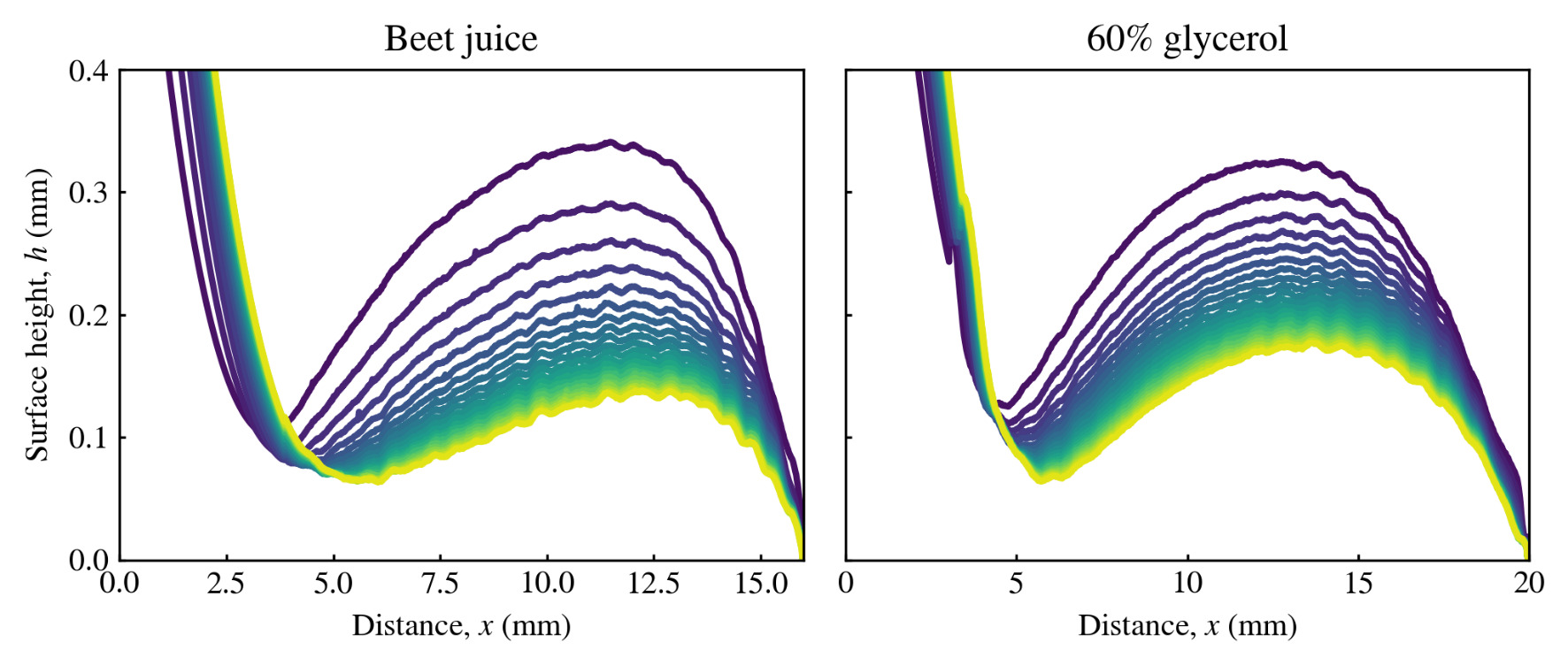}
    \caption{
    \hl{Surface profile evolution of beet juice and 60\% glycerol-water mixture.}
    }
    \label{fig:marangoni}
\end{figure}

\hl{The asymmetric surface profile may be due to the Marangoni effect. Although we did not test this explicitly, as this is not a major point of this work, we observed more symmetric surface profiles in glycerol-water mixtures, compared to beet juice. 
Figure~\ref{fig:marangoni} compares the surface profiles of beet juice and 60\% glycerol-water.}

\newpage

\section{Dimple time data}\label{app:all-dimple-time-data}

\begin{figure}[ht]
    \centering
    \includegraphics[width=.8\linewidth]{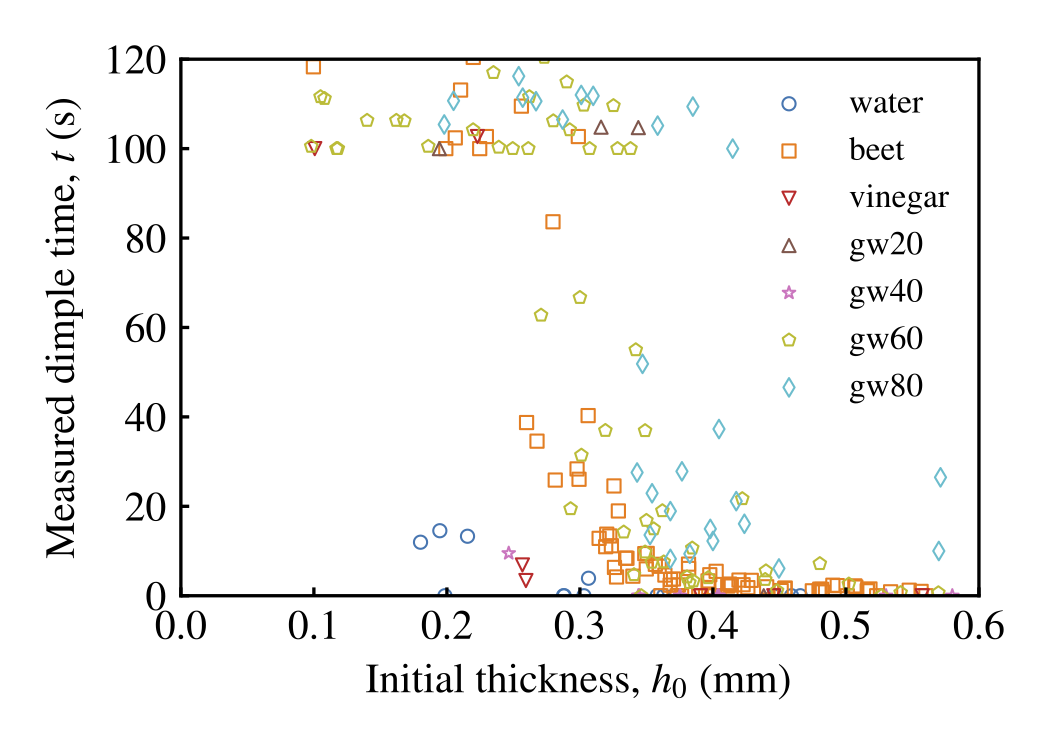}
    \caption{
    Alternative way to look at the $t_{\mathrm{dimple}}$ data. Most of the times are either 0 or above 100 s, but a clear separation in the initial thickness $h_0$ can be identified, highlighting that long dimples are only observed for small $h_0$.
    }
    \label{fig:all-dimple-time-data}
\end{figure}

\bibliographystyle{apsrev4-2}

%

\end{document}